\documentclass[journal]{IEEEtran}
\usepackage[utf8]{inputenc}
\usepackage{graphicx}
\usepackage{subfigure}
\usepackage{textcomp}
\usepackage{amsmath}
\usepackage{amsfonts}
\usepackage{xspace}
\usepackage{cite}
\usepackage{url}
\usepackage{enumerate}
\usepackage{epstopdf}
\usepackage{color}
\usepackage{color,soul}

\usepackage[]{todonotes}

\graphicspath{{figures/},{figurepdfs/}}
\newcommand{\Lower}[1]{\smash{\lower 1.5ex \hbox{#1}}}

\def\putbox#1#2#3#4{\makebox[0in][l]{\makebox[#1][l]{}\raisebox{\baselineskip}[0in][0in]{\raisebox{#2}[0in][0in]{\scalebox{#3}{#4}}}}}
\def\rightbox#1{\makebox[0in][r]{#1}}
\def\centbox#1{\makebox[0in]{#1}}
\def\topbox#1{\raisebox{-0.60\baselineskip}[0in][0in]{#1}}
\def\midbox#1{\raisebox{-0.20\baselineskip}[0in][0in]{#1}}

\begin{document}
\title{1/$f$ Noise Reduction using In-Pixel Chopping in CMOS Image Sensor}
\author{Kapil Jainwal and Mukul Sarkar, \textit{Member IEEE}
\thanks{Manuscript received June X, XXXX.}
\thanks{Authors are with the Electrical Engineering Department, Indian Institute of Technology Delhi, 110016 New Delhi, India. (e-mail: KJ: kapiljainwal@gmail.com; MS: msarkar@ee.iitd.ac.in) 
}% <-this % stops a space
}

\markboth{Journal of \LaTeX\ Class Files,~Vol.~13, No.~9, January~2016}%
{Shell \MakeLowercase{\textit{et al.}}: Bare Demo of IEEEtran.cls for Journals}
      
\maketitle
% Abstract
\begin{abstract}
\label{sec:ABSTRCT}
In this paper, an in-pixel chopping technique to reduce the low-frequency or 1/\textit{f} noise of the {\color{black}source follower (SF)
transistor in active pixel sensor (APS) is presented}. % One of the major reasons responsible for the low-frequency noise is the trap state randomization of defects
% present at the gate Si-SiO$_\textbf{2}$ interface of MOSFET.
The SF low-frequency noise is modulated at higher frequencies through chopping, implemented inside the pixel, and in later stage eliminated using low-pass filtering. To implement the chopping, {\color{black} the conventional 3T APS architecture is modified, with only one additional transistor of minimum size per pixel.}
Reduction in the noise also enhances the dynamic range {\color{black}(DR)} of the image sensor. 
The test circuit is fabricated in UMC 0.18 \textbf{$\mu$}m standard CMOS technology. The measured results show a reduction of 1/\textit{f} noise by approximately 22 dB for 50 MHz chopping frequency {\color{black}(\textit{$\textbf{f}_\textbf{ch}$})}. 
% and {\color{black}increases the SNDR by approximately 10 dB. } %with XX dB} dynamic range. 
\end{abstract}
\section{Introduction}
%\vspace{-0.05 cm}
\label{sec:INTRO}
%Apart from applications like the integration of high-performance miniature circuits, CMOS technology is now becoming a popular choice for sensors design.  In the present era, CMOS image sensors are being extensively used in the digital image sensing systems.  % CMOS image sensors are now popular choices for use in image sensing systems.
\IEEEPARstart In the present era, CMOS image sensors are being extensively used in digital imaging systems. 
{\color{black}High dynamic range (DR)} is one of the primary performance defining parameters for a CMOS image sensor. The dynamic range is limited by the output swing and high noise. 
%High noise and limited output swing are the bottlenecks for a high dynamic range. 
The primary sources of noise in an active pixel sensor (APS) of a CMOS imager are the thermal noise from the switches and the low-frequency noise from the source follower (SF). The thermal noise from reset switch of the pixel can efficiently be reduced using correlated double sampling (CDS). The low-frequency or 1/$f$ noise of the SF remains as a major source of noise in an active pixel. 
%As the 1/$f$ noise is inversely proportional to the area of the transistor, it increases significantly when the technology shrinks down to deep sub-micron level.

The 1/$f$ noise results from the random telegraph signal (RTS) which causes the discrete fluctuation of the conducting current or the threshold voltage and eventually produces the blinking output behavior of the pixel. The image quality is severely affected by this random behavior as the blinking is very visible to human eyes. % in most imaging applications.
%CDS can also reduce 1/$f$ noise, however, the reduction depends on the sampling frequency of the CDS and is only effective when the sampled noise components are correlated \cite{paper:Enz_CDS_chop, paper:KANSY_CDS , paper:wangTHESIS}. %,paper:MICRO_chop_CDS}. 
%
%Bloom and Nemirovsky \cite{paper:SWITCH_Bloom91} have introduced the 1/$f$ noise reduction technique based on cycling a MOS transistor between strong inversion and accumulation region.
%
% Later the switching technique is further investigated, modeled and verified by several researchers \cite{paper:SWITCH_Gierkink99,paper:SWITCH_Klumperink2000,paper:SWITCH_VEN_DER_2000}. This technique is commonly used in many systems like phase locked loop (PLL), in which switching is used to reduce the phase noise of a voltage controlled oscillator (VCO). In \cite{paper:Mypaper} the switching of MOS transistor has been employed to reduce the 1/f noise in APS using shared source followers among the pixels.  
%
To reduce the $1/f$ noise a pMOS SF transistor is used in~\cite{ISSCC2011}. The use of pMOS transistor reduces the fill-factor of the pixel. {\color{black} In~\cite{ISSCC2012} a buried channel SF while, in ~\cite{TED16} a thin oxide pMOS transistor is used to reduce the $1/f$ noise.}
%In~\cite{ISSCC2012} a buried channel SF is used to reduce the $1/f$ noise. In~\cite{TED16} a thin oxide pMOS transistor is used to reduce the $1/f$ noise. 
The $1/f$ noise reduction technique in~\cite{ISSCC2012} and~\cite{TED16} needs process {\color{black} modifications} and thus, would increase the cost. The cycling of a MOS transistor between strong inversion and accumulation also reduces $1/f$ noise~\cite{paper:SWITCH_Bloom91, paper:SWITCH_Gierkink99, paper:Mypaper2}.      
 \begin{figure}
 \centering
    \includegraphics[scale =0.7]{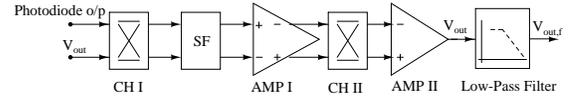}
  %\vspace{-0.3 cm}
 \caption{{\color{black}Block diagram for in-pixel chopping.  }}
  %\vspace{-0.4 cm}
 \label{fig:BD_INPXL} 
                                 
 \end{figure}

 \begin{figure*}[ht!]
 \centering
 \def\putbox#1#2#3#4{\makebox[0in][l]{\makebox[#1][l]{}\raisebox{\baselineskip}[0in][0in]{\raisebox{#2}[0in][0in]{\scalebox{#3}{#4}}}}}
 \def\rightbox#1{\makebox[0in][r]{#1}}
 \def\centbox#1{\makebox[0in]{#1}}
 \def\topbox#1{\raisebox{-0.60\baselineskip}[0in][0in]{#1}}
 \def\midbox#1{\raisebox{-0.20\baselineskip}[0in][0in]{#1}}
 \includegraphics[scale = 1]{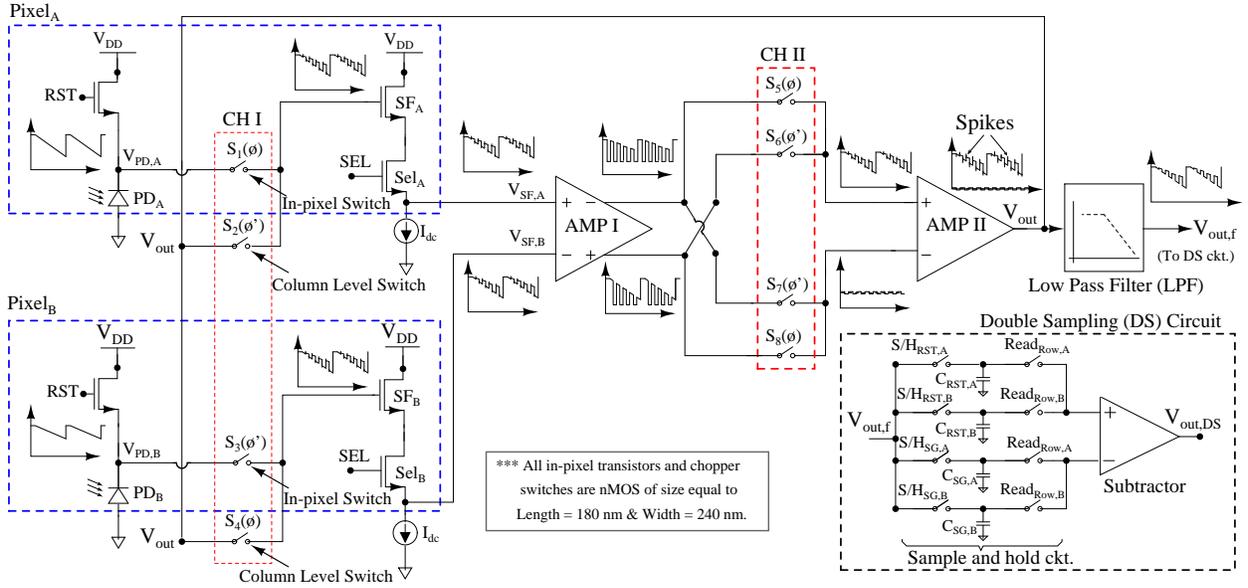}
 %    \putbox{-4.6 in}{2.07 in}{1}{{Sel}$_A$}%
%    \putbox{-4.6 in}{0.55 in}{1}{{Sel}$_B$}%
 %\vspace{-0.3 cm}                                        
 \caption {{\color{black}Circuit diagram of in-pixel chopping - excluding  Pixel{$_A$} and Pixel{$_B$}, other blocks are placed in column level readout circuitry}}
  %\vspace{-0.4 cm}                                        
 \label{Ckt_inPxl_Chop}                                                      \end{figure*}
 Chopping  \cite{paper:CHOP4} is used for $1/f$ noise reduction, in which the low-frequency noise is modulated to the chopping frequency ($f_{ch}$) far beyond the frequency band of interest. Chopping needs extra switches and would hamper the fill-factor of the pixel and thus has never been used in a pixel. In this work, a novel technique is presented to implement chopping inside a {\color{black}conventional 3T pixel}. The basic building block of the proposed in-pixel chopping is shown in Fig. \ref{fig:BD_INPXL}. The photodiode (PD) output signal is modulated to the chopping frequency ($f_{ch}$) using the first chopper (CH I) before being buffered by the SF. The output of the SF is fed to the input of the amplifier stage I (AMP I). The output of the AMP I is composed of the amplified modulated PD signal, amplified low-frequency noise from the SF, and the noise and output offset of the AMP I. The output of the AMP I is chopped again using the second chopper (CH II). The AMP I and the CH II are placed in the column and does not affect the fill-factor of the pixel. The CH II demodulates the PD signal to its original baseband frequency whereas, modulates the low-frequency noise of the SF and AMP I, and amplifier offset voltage to $f_{ch}$. After CH II the signals are further amplified by the amplifier stage II (AMP II). The output of the AMP II is fed back to the other input of CH I to complete the closed loop unity gain feedback configuration. A low-pass filter (LPF) followed by CH II suppresses the up-modulated offset and 1/$f$ noise and also blocks the spikes in the output at the frequency $f_{ch}$, generated due to chopping action.
      
  The modified pixel read-out helps in achieving the functionality as well as a reduction in the low-frequency noise.  
  {\color{black} An additional minimum sized transistor is used in-pixel as compared to 3T pixel. To compensate the fill-factor a minimum sized SF is used, the noise of which is reduced using in-pixel chopping.
  %The fill-factor of the pixel is not affected much as only one minimum sized transistor and a feedback signal line, are additionally used per pixel, as compared to a standard 3T APS. To compensate the area penalty caused by an additional switch, minimum sized SF is used, the noise of which is reduced using chopping.
} The low-frequency noise power reduces by 22 dB, as compared to a conventional 3T APS without chopping. The rest of the paper is organized as follows: in-pixel chopping implementation is described in section \ref{sec:SYS}, simulation and experimental results are presented in section \ref{MSR_SIM}, and the paper is concluded in section \ref{CONCLUSION} .
%The rest of the paper is organized as follows: The in-pixel chopping and the low-frequency noise reduction mechanism is described in the Section \ref{sec:SYS}. Section \ref{READ_OUT} describes the read-out of the proposed imager. The simulation and experimental results are presented in section \ref{MSR_SIM}. The paper is concluded in section \ref{CONCLUSION}.
% Enhancement in the DR from the in-pixel chopping is described in Section \ref{subsec:DR_ENHNCE}}. 
%
%
%\vspace{-0.2 cm}
 \section{System Design}
 \label{sec:SYS}
{\color{black}

 The circuit diagram for the in-pixel chopping in active pixel sensor of a CMOS imager is shown in {\color{black}Fig. \ref{Ckt_inPxl_Chop}}.   
  {\color{black}The two pixels A and B consist of photodiodes PD$_A$, PD$_B$,  chopper switches S$_1$, S$_3$ (of CH I), select switches Sel$_A$, Sel$_B$, and source followers SF$_A$, SF$_B$. %In-pixel circuit is shown in the box in Fig. \ref{Ckt_inPxl_Chop}. 
  The remaining two switches S$_2$, S$_4$ of CH I, AMP I, CH II (switches S$_5$-S$_8$), AMP II, and a low-pass filter are placed in column level circuits. At the onset, the photodiodes of Pixel$_{A}$ and Pixel$_{B}$ are reset to $V_{rst}$ using RST switch. Light is then integrated on the photodiodes. After the integration time, the photodiode output signals V$_{PD,A}$ and V$_{PD,B}$ are modulated to chopping frequency $f_{ch}$ using switches S$_1$-S$_4$ of CH I. The non-overlapping clock signals $\phi$ and ${\phi}$' run at the fundamental chopping frequency $f_{ch}$.  
During readout, Pixel$_{A}$ and Pixel$_{B}$ are selected together using the select signal SEL at the input of switches Sel$_A$ and Sel$_B$. The row decoder of the image sensor selects two rows at a time for simultaneous selection of two adjacent pixels in the column. The SF output of Pixel$_{A}$ and Pixel$_{B}$ are amplified using AMP I. The AMP I is realized using a folded cascode differential input differential output amplifier.

 The clock signal $\phi$ turns {\color{black}the switches S$_1$, S$_4$} and S$_5$, S$_8$  ON  for a  time interval $t_1$ and the clock signal ${\phi}$' turns {\color{black}the switches S$_2$, S$_3$} and S$_6$, S$_7$ ON, for $t_2$  ($t_1$ and $t_2$ are non-overlapping and equal time intervals). After modulation of the photodiode signals V$_{PD,A}$ and V$_{PD,B}$ through CH I %(S$_1$-S$_4$) 
  and buffered by the SF, the input of the AMP I can be given as
\begin{equation}
\begin{aligned}
%%\vspace{- 0.4 cm}
V_{SF,A} = V_{PD,A}(t_1) + V_{out}(t_2) + N_{sf,A} , \\
V_{SF,B} = V_{PD,B}(t_2) + V_{out}(t_1) + N_{sf,B} ,
\end{aligned}
%\vspace{-0.5 cm}
\end{equation}
where N$_{sf,A}$ and N$_{sf,B}$ are the low-frequency noise from SF$_A$ and SF$_B$, respectively. The notation of V$_{PD,A}(t_1)$ is chosen to denote the signal V$_{PD,A}$ during $t_1$ time interval and also applicable to similar terms.  {\color{black}In next stage the output of AMP I is chopped using CH II, which demodulates the photodiode signal to the baseband and modulate the offset and low-frequency noise to $f_{ch}$. CH II consists of switches S$_5$-S$_8$ operated on same non-overlapping clocks $\phi$ and ${\phi}$'. The differential output of the CH II is amplified by single-ended difference amplifier AMP II. The output of AMP II is fed back to the Pixel$_{A}$ and Pixel$_{B}$ to close the loop. Thus, AMP I and II both are required to form a closed loop unity gain system.   
    
If AMP I and AMP II has a voltage gain of $A_1$ and $A_2$, offset of $V_{of1}$ and $V_{of2}$, low-frequency noise of $N_{Am1}$ and $N_{Am2}$, respectively, the output signal V$_{out}$ is expressed as}
%%\vspace{-0.1 cm}
\begin{equation}
\label{vouta}
\centering
\begin{aligned}
V_{out} = [V_{PD,A}(t_1)] + V_{PD,B}(t_2)] 
+ [N_{sf,A}(t_1) - N_{sf,A}(t_2)] \\ - [N_{sf,B}(t_1) - N_{sf,B}(t_2)] 
+ \frac{N_{Am1}(t_1) - N_{Am1}(t_2)}{A_1} \\ + \frac{V_{of1}(t_1) - V_{of1}(t_2)}{A_1} 
+ \frac{N_2 + V_{off2}}{2A_1A_2}. 
\end{aligned}
\end{equation}

    \begin{figure*} [ht!]
    \centering
    \includegraphics[scale =0.59]{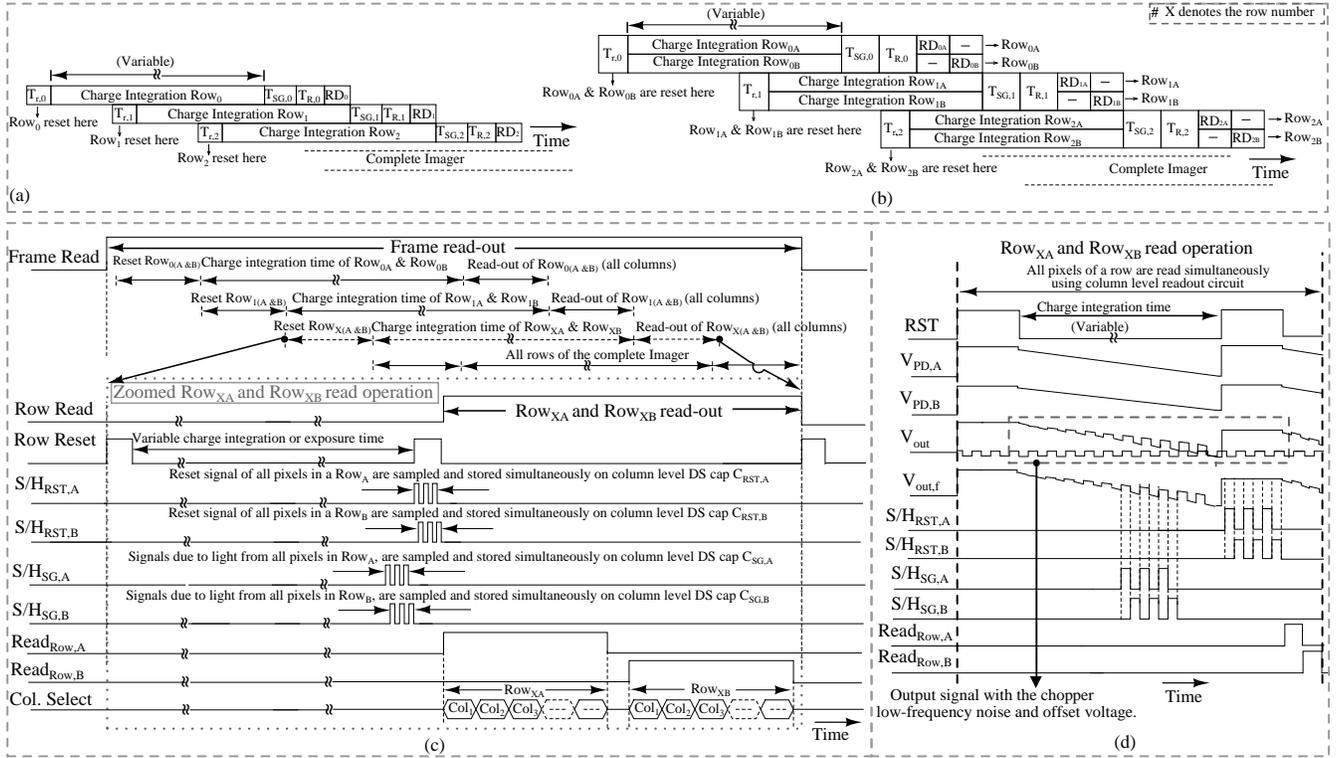}
           %\vspace{-0.3 cm}                                        %\includegraphics[scale = 0.6]{FIG3_TIMING3}
    \caption {{\color{black} (a) Conventional rolling shutter (b) proposed readout. Pixels of Row$_{XA}$ and Row$_{XB}$ are reset during the time interval T$_{r,X}$. After a variable charge integration time, the output signal of all pixels of Row$_{XA}$ and Row$_{XB}$ are sampled and stored on column level capacitors during T$_{SG,X}$. Row$_{XA}$ and Row$_{XB}$ are reset again and sampled and stored on column level capacitors during T$_{R,X}$. Row$_{XA}$ and Row$_{XB}$ are read-out during RD$_{XA}$ and RD$_{XB}$, respectively, (c) Timing diagram of a frame read-out for the imager based on proposed technique, and (d) Timing diagram of Row$_{XA}$ and  Row$_{XB}$ read-out.}}
     %\vspace{-0.4 cm}                                        
    \label{TIMING} 
    \end{figure*} 
  
If the small signal voltage gain values $A_1$ and $A_2$ are very high, then only the photodiode signals V$_{PD,A}$ and V$_{PD,B}$ along with 1$/f$ noise from the source followers dominate and the output signal  V$_{out}$ can be simplified as 	 
% %\vspace{-0.5 cm}
\begin{equation}
\label{voutb}
\begin{aligned}
 V_{out} \approx [V_{PD,A}(t_1) + V_{PD,B}(t_2)] ~~~~~~~~~~~
\\ +~[N_{sf,A}(t_1) - N_{sf,A}(t_2)]- [N_{sf,B}(t_1) - N_{sf,B}(t_2)]. 
\end{aligned}
%%\vspace{-0.05 cm}
\end{equation}

%{\color{black}The expression derived as  (\ref{vouta}) and (\ref{voutb}) are true when the noise components are correlated, which is true for lower frequency noise and when the chopping frequency is much higher than the low-frequency noise corner frequency.  While deriving these equations, the noise pairs like N$_{sf,A}$(t$_1$) and N$_{sf,A}$(t$_2$) are assumed to be correlated because of the higher chopping frequency is used for the measurements \cite{paper:SWITCH_VEN_DER_2000,paper:Mypaper1,paper:Mypaper2}.} 

{\color{black} In (2) and (3), it is assumed that the chopping frequency is much higher than the low-frequency noise corner frequency and the noise pairs like N$_{sf,A}$(t$_1$) and N$_{sf,A}$(t$_2$) are correlated due to high $f_{ch}$~\cite{paper:SWITCH_Gierkink99,paper:Mypaper2}.

%In (2) and (3), it is assumed that the chopping frequency is much higher than the low-frequency noise corner frequency and the noise pairs like N$_{sf,A}$(t$_1$) and N$_{sf,A}$(t$_2$) are correlated due to high chopping frequency~\cite{paper:SWITCH_VEN_DER_2000,paper:Mypaper1,paper:Mypaper2}.
}

 To suppress the overall input referred noise and offset at the output (from the amplifier stages), a two-stage high gain amplifier is designed for column readout circuit. The amplifier system has 20-MHz unity gain-bandwidth with a phase margin of 65$^0$ and maximum power consumption of {200 $\mu$W}. The first stage of the opamp is a differential input/differential output folded cascode amplifier (AMP I with small signal voltage gain of 68 dB), which is followed by a difference amplifier with a single-ended output (AMP II with small signal voltage gain of 40 dB) to achieve an overall small signal voltage gain of 108 dB.          
%
%
%
%   
%
% \begin{figure*}
%    %\includegraphics[width=12cm,height=5.5cm]{PXL_4T_CHOP} 
%   \centering
%   \includegraphics[scale =0.57]{FIG3_TIMINGN.eps}
%   %\includegraphics[scale = 0.6]{FIG3_TIMING3}
%                                     
%   \caption {{\color{black}Timing diagram of the pixel readout with chopping and CDS operation.}}
%   \label{TIMING} 
%   
% \end{figure*}
  
The output signal V$_{out}$ is continuous and composed of Pixel$_{A}$ output for time duration $t_1$ and Pixel$_{B}$ output for time duration $t_2$, periodically. The switches used for chopping introduces ripples at the output. These ripples are generated due to clock feed-through of the overlapping capacitance present between drain and gate of the switching transistors. A switched capacitor low-pass filter is used to block the ripples present in the output signal \cite{paper:Filter1}. 

%{\color{black}The 1/$f$ noise power and offset at low frequencies are very less as modulated to $f_{ch}$. The output of the AMP II is filtered using a switched capacitor low-pass filter (LPF) \cite{paper:Filter1}. The LPF filters out the modulated 1/$f$ noise and offset voltage. The switches used for chopping introduces ripples at the output. These ripples are generated due to clock feed-through of the overlapping capacitance present between drain and gate of the switching transistors \cite{paper:Clock_feed_thourgh1}. The LPF also filters out the ripples present in the output signal. The output of the LPF now only consists of the photodiode signals, which can be sampled at separate time instances to differentiate the output signal from Pixel$_{A}$ and Pixel$_{B}$. The output of the LPF is fed to the double sampling circuit to eliminate the pixel level fixed pattern noise (FPN). % A conventional CDS circuit is used in this design.  

% The proposed in-pixel chopping method effectively reduces the low-frequency noise without adding any extra switches in the conventional 4T pixel. Thus, the fill-factor of the pixel is unaffected. %The 1/$f$ noise reduction is obtained by changing the readout method as compared to a conventional pixel along with additional column level chopping circuitry. 
As the dynamic range of a conventional APS is limited by the noise level, the technique also enhances the DR of the pixel. The photodiode signal {\color{black}gets} buffered through a chopper amplifier including SF, high gain amplifier stage I and II (configured in closed loop with unity gain) and the final output is fed back to one of the inputs of first chopper CH I. Hence, the continuous output of the closed-loop chopper amplifier is virtually short with the photodiode output node. The high gain of the amplifier (108 dB) make the output follow the photodiode node linearly for a wide range of light integration, increasing the the output swing and dynamic range.
%As the gain of the amplifier is very high (108 dB) the output node follows the photodiode node linearly for the wider range of light integration as compared to the case of source follower buffer. This eventually increases the swing of the output voltage and hence, the dynamic range of the image sensor.  In the proposed architecture the dynamic range is enhanced due to both reduction in the low-frequency noise and an increase in the SF output swing.    
%
%\begin{figure}
%\includegraphics[scale = 0.4]{FIG4_CHOP_SIM2} 
  % \includegraphics[scale=1.3]{PXL_4T_CHOP}
%\centering
%%\vspace{-0.3 cm}
%\caption {{\color{black}Post layout $1/f$ noise PSD Simulation (PSS~+~PNoise) results (using Cadence simulator tool - Spectre) without and with in-pixel chopping (varying $f_{ch}$ from 800 kHz to 5 MHz). Input signal fundamental frequency is 50 kHz.}}
%%\vspace{-0.2 cm} 
%\label{SIM_NOISE}
%\end{figure}
%
%
{\color{black}

The read-out timing diagram for the in-pixel chopping operation and the output waveforms of the critical nodes are shown in Fig.~\ref{TIMING}(a)-(d). The read-out is based on conventional rolling shutter mode as in 3T pixel. The conventional and proposed read-out modes are shown in Fig. \ref{TIMING}(a) and (b), respectively. %The readout is shown in Fig. \ref{TIMING} (d), in which a part of the charge integration time of rows are overlapped, is similar to the conventional rolling shutter.
However, in the proposed architecture instead of a single row, two adjacent rows, Row$ _{XA} $ and Row$ _{XB} $ (X is used to denote the row number, for example, Row$_{1A}$ and Row$_{1B}$) are selected together for readout. 
Charge integration on photodiode, charge to voltage conversion, chopping/de-chopping of the photodiode signal, signal due to light/reset level sample and hold, double sampling (DS) and low-frequency noise filtering are carried out on the pixel pairs (i.e. Pixel$_{ 0A }$-Pixel$_{ 0B }$, Pixel$_{ 1A }$-Pixel$_{ 1B }$, Pixel$_{ 2A }$-Pixel$_{ 2B }$.....) for Row$_{XA }$ and Row$_{XB}$  together. 
%These operations on each pair of the pixels of the selected rows (Row$ _{XA} $ and Row$ _{XB} $) are carried out in parallel using column level circuit. 
The timing diagram of the in-pixel chopping architecture is shown in Fig. \ref{TIMING}(c) and (d).   \par 
The double sampling circuit is modified to sample and hold the reset and signal of the pixel pair of adjacent rows, as shown in Fig. \ref{Ckt_inPxl_Chop}. The reset signal of Pixel$_A$ and Pixel$_B$ of 
Row$_{XA}$ and Row$_{XB}$ are sampled on capacitors C$_{RST,A}$ and C$_{RST,B}$, while the signal after a variable integration period is sampled on capacitors C$_{SG,A}$ and C$_{SG,B}$, respectively.
%During the reset phase Row$_{XA}$ and Row$_{XB}$ are reset and after a variable charge integration or exposure time the signals from Pixel$_A$ and Pixel$_B$ are sampled on sampling capacitors  C$_{SG,A}$ and C$_{SG,B}$, respectively. {\color{black} Then, Row$_{XA}$ and Row$_{XB}$ are reset again and the reset levels of Pixel$_A$ and Pixel$_B$ are sampled on the reset capacitors C$_{RST,A}$ and C$_{RST,B}$. }
%Then, Row$_{XA}$ and Row$_{XB}$ are reset again and the reset levels of pixel$_A$ and pixel$_B$ of Row$_{XA}$ and Row$_{XA}$, respectively are sampled on the reset capacitors C$_{RST,A}$ and C$_{RST,B}$, respectively of the double sampling circuit.
 Switches S/H$ _{RST,A} $ and S/H$ _{RST,B} $ are ON for repetitive and non-overlapping time intervals $t_{ 1 }$ and $t_{ 2 }$, respectively,  sampling the reset levels, while, switch S/H$ _{SG,A} $ and S/H$ _{SG,B} $ are ON similarly, sampling the output signals. 
%This sampling is performed on all pixel pairs of the Row$_A$ and Row$_B$, simultaneously using column level sample and hold circuit.
%After storage of reset level and output signals, delta differential sampling (DDS) is performed to cancel the pixel level fixed pattern noise (FPN) for Row${_A}$ by turning the switch Read$_{Row,A}$ ON.
The output of all pixels of Row$_A$ and Row${_B}$ are then, sequentially read by turning the switch Read$_{Row,A}$ and Read$_{Row,B}$ ON, respectively.}
% After reading the Row$_A$, the output of Row${_B}$ are read by turning the switch Read$_{Row,B}$ ON. }    
%\par
%However, during the double sampling, the sampled $k$T/C noise ($k$ is Boltzmann constant, T is temperature, and C is capacitance) components of the reset switch are non-correlated as they come from two different reset phases, Thus, instead elimination, DDS eventually increases the thermal noise, which is a well-known limitation of 3T pixel readout.}
%
%
%
\begin{figure}
	\includegraphics[scale = 1.2 ]{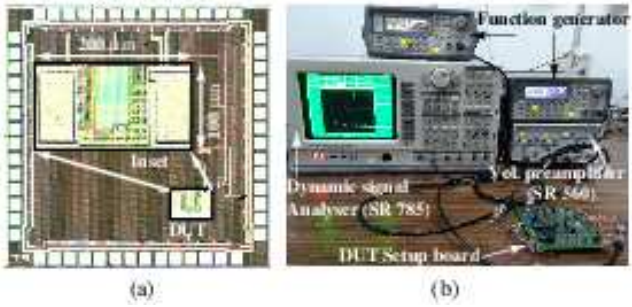} 
	\centering 
	%\vspace{-0.3 cm}
	\caption {{\color{black}(a) Chip micro-photograph, (b) Measurement setup}}
	%%\vspace{-0.4 cm} 
	\label{DUT}
\end{figure} 
\begin{figure*}[t]
\includegraphics[scale = 0.262]{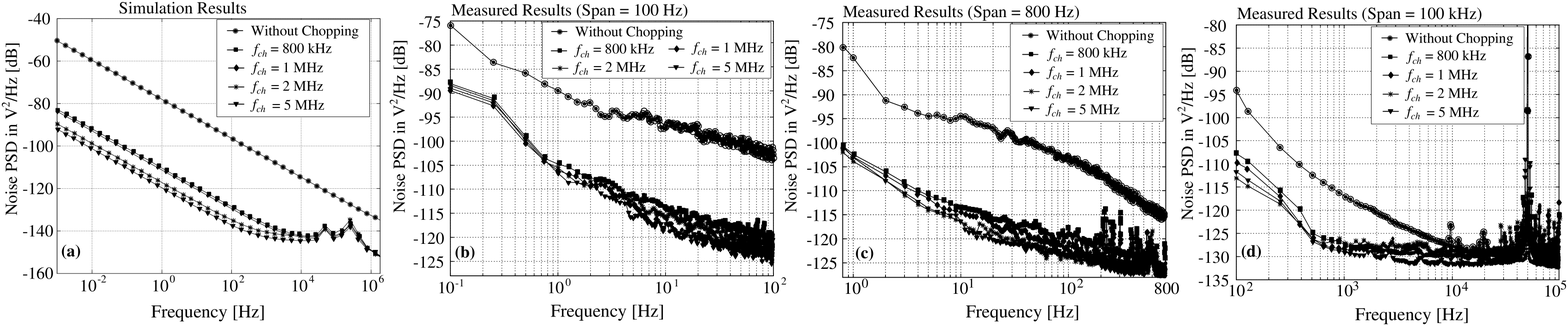} 
\centering 
%%\vspace{-0.4 cm} 
\caption{{\color{black}(a) Post layout $1/f$ noise PSD Simulation (PSS~+~PNoise) results (using Cadence simulator tool - Spectre) with and without in-pixel chopping, ~~~~~~
Measured low-frequency noise PSD (Fig. (b), (c), and (d)) for variable chopping frequencies $f_{ch}$ (from 800 kHz to 5 MHz) and sampling frequency span of (b) 100 Hz, (c) 800 Hz, and  (d) 100 kHz. [The input signal fundamental frequency for all results is 50 kHz].}}
%%\vspace{-0.4 cm} 
\label{MSR_NOISE}
\end{figure*}
\section{Simulation and Measurement Results}
\label{MSR_SIM}
The post-layout noise PSD simulation results, shown in {\color{black}Fig. \ref{MSR_NOISE} (a)} are generated by the periodic steady state (PSS) and Pnoise analysis in Cadence IC-615 using Star-Hspice 49 models for UMC 0.18 $\mu$m process. 
%The MOSFET model used is Star-Hspice level 49 (BSIM3V3.2) with UMC 0.18 $\mu$m Mixed-mode/RFCMOS 1.8V 1P6M P-sub twin-well CMOS salicide process. 
{\color{black} The output noise power (integrated in the frequency band from 1 Hz to 10 kHz) without chopping is -65.01 dB. Whereas, with chopping the integrated noise power reduces to -93.55 dB, -94.28 dB, -99.91 dB, and -100.27 dB for $f_{ch}$ equal to 800 kHz, 1 MHz, 2 MHz, and 5 MHz, respectively for an input signal of 50 kHz. 
%The noise results are shown for an input signal of 50 kHz fundamental frequency,  chopped at varying frequency from 800 kHz to 5 MHz. The PSS~+~Pnoise analysis simulation results show the output noise power (integrated noise power in the frequency band from 1 Hz to 10 kHz) without chopping is -65.01 dB. Whereas, with chopping the integrated noise power reduces to -93.55 dB, -94.28 dB, -99.91 dB, and -100.27 dB for $f_{ch}$ equal to 800 kHz, 1 MHz, 2 MHz, and 5 MHz, respectively.} % This shows the noise reduction of 20.16 dB, 20.41 dB, 21.62 dB, and 22.31 dB for chopping frequencies of 800 kHz, 1 MHz, 2 MHz, and 5 MHz, respectively.
}
%For the noise PSD simulation results shown in Fig. \ref{SIM_NOISE}, the value of $f_s$ is chosen as 10 kHz because the noise reduction of approximately 25 dB at this switching frequency is most closer to the measured results given in upcoming part in this section.   

The in-pixel chopping architecture is fabricated in 0.18 $\mu$m 1P6M  standard CMOS process and the microchip photograph is shown in Fig. \ref{DUT}(a).  
%\todo[inline]{a) give the make of the photodiode. n-well/p-sub is what you must have used. If PD has not been used, specifically how the PD output is generated.  b) The microchip photograph should show where the pixel A and B are. c) why are you not using the die photograph? d) what is the pitch, the imager projection will need pitch.\\Ans: Edited the caption and } 
{\color{black} The design under test (DUT) consists of two-pixels with the chopper amplifier. The test pixel is without photodiode and the input signal for the SF is a replica of photodiode output signal, generated from a function generator during measurements.} The measurement setup is shown in {\color{black}Fig. \ref{DUT}(b) and is similar to that reported in \cite{paper:Mypaper2}.} The measured noise PSD of the DUT for varying frequency span of 100 Hz, 800 Hz, and 100 kHz is shown in {\color{black}Fig. \ref{MSR_NOISE} (b), (c), and (d)}, respectively. For each span, the noise PSD of the pixel is shown, without chopping and with in-pixel chopping for $f_{ch}$ equal to 800 kHz, 1 MHz, 2 MHz, and 5 MHz. To improve the measurement accuracy, each noise PSD curve is plotted after taking an RMS average of 1000 measured samples.

The in-pixel chopping reduces the $1/f$ noise for all $f_{ch}$ greater than twice the fundamental frequency of the input signal (50 kHz) during measurements.
%Results with variable $f_{ch}$ are given to show that the technique works properly for all chopping frequencies greater than the fundamental frequency of the input signal which is kept at 500 Hz during measurements.
%Chosen sampling frequency is suitable to compare the proposed circuit with CMOS image sensor and chopping frequencies have been kept much higher than sampling frequency as mentioned.
{\color{black} The 1/$f$ corner frequency which is around 10 kHz without chopping is shifted to below 1 kHz (around at 800 Hz) with chopping. The integrated noise power from 1 Hz to 10 kHz without chopping is -75.74 dB  (163.23 $\mu$V$_{RMS}$), whereas with chopping the integrated noise power reduces to -95.90 dB  (16.023 $\mu$V$_{RMS}$), -96.15 dB (15.53 $\mu$V$_{RMS}$), -97.36 dB (13.55 $\mu$V$_{RMS}$), and -98.05 dB (12.51 $\mu$V$_{RMS}$) for 800 kHz, 1 MHz, 2 MHz, and 5 MHz, respectively. This shows the noise reduction of 20.16 dB, 20.41 dB, 21.62 dB, and 22.31 dB for chopping frequencies of 800 kHz, 1 MHz, 2 MHz, and 5 MHz, respectively. }

{\color{black}The low-frequency noise reduction using in-pixel chopping is compared with recently reported noise performances for CMOS imager in Table \ref{TAB:Table2}. The integrated RMS noise for 1 Hz to 10 kHz frequency band, at the output of the SF without chopping is 163.23 $\mu$V$_{RMS}$, which gets reduced to 12.5 $\mu$V$_{RMS}$ using in-pixel chopping for 5 MHz chopping frequency.} As observed from the Table \ref{TAB:Table2} the proposed work results in the lowest noise as compared to other methods. The use of buried channel nMOS and thin oxide pMOS SF needs process modifications. However, the proposed method uses the conventional fabrication process. Further use of pMOS SF reduces the pixel fill-factor in [6]. The proposed in-pixel chopping uses an nMOS SF thus, does not compromise much with the fill-factor of the pixel.}
% Further, the swing of the output increases by the noise reduction achieved through the in-pixel chopping and fill-factor  techniques. This could helps in enhancing the overall dynamic range of the CMOS image sensor.
% {\color{black}Further, the swing of the output increases by the noise reduction achieved through the in-pixel chopping and fill-factor  techniques. The non-linearity or distortion in the output signal also decreases from virtual short configuration from the closed-loop chopper amplifier. The measured SNDR at the pixel output increases from 70 dB to 80.5 dB after employing the in-pixel chopping technique. This helps in enhancing the overall dynamic range of the CMOS image sensor.}    
        %%\vspace{-0.2 cm}
        \begin{table}
          \centering
          \caption{Performance comparison of the noise in CMOS Imagers}
             \label{tab:Table 2}
            \begin{tabular}{p{3.7 cm}|p{1.9 cm}|p{1.6 cm}}
            \centering 
           \textbf{Technique}  & \centering  \textbf{Noise}\
           [\textbf{$\mu$}\textbf{V}$_{\textbf{RMS}}$] &  \textbf{Reference}  \\
            \hline
            \hline
           pMOS in-pixel Amplifier   & \centering  258 & ISSCC'11 \cite{ISSCC2011}       
                  \\
                \hline
                Burried Channel nMOS SF, Multiple Sampling with SSADC  & \centering  31.5 & ISSCC'12 \cite{ISSCC2012}  \\
                \hline
                Thin Oxide pMOS SF & \centering  74 & TED'16 \cite{TED16}\\
%                \hline
%                Column-Parallel Digital Correlated Multiple Sampling & \centering  127 & Sensors Jour. (2012) \\
                \hline
                In-Pixel Chopping  & \centering  {\color{black} 12.5}  & This Work \\ 
               \hline
           \hline
          \end{tabular}
      %%\vspace{-0.5 cm}
      \label{TAB:Table2}
          \end{table}
\section{Conclusion}
\label{CONCLUSION}
{\color{black}The in-pixel chopping is applied to the conventional 3T pixel which reduces the low-frequency noise of the source follower. 
%The technique is implemented by adding a minimum size extra switch inside per pixel. 
A reduction in the integrated noise power of 22 dB with 5 MHz chopping frequency, is obtained. The reduction in the low-frequency noise improves the dynamic range of the image sensor and hence, can be used to improve the quality of the image. The reduction in the fill-factor due to an extra in-pixel switch can partially be compensated by choosing a minimum size source follower. {\color{black}The noise of the minimum size source follower is reduced using in-pixel chopping. } }} %Usually, a larger source follower transistor is to be used for a reduced 1/$f$ noise. The reduction of the source follower noise using in-pixel chopping will allow using a smaller source follower, improving the effective fill-factor. }

\end{document}